\documentclass[twocolumn,showpacs,floatfix,superscriptaddress]{revtex4}
\usepackage{graphicx}
\usepackage{bm}

\begin{document}

\title{Finite temperature effects on the collapse of trapped Bose-Fermi mixtures}
\author{Xia-Ji Liu}
\affiliation{\ LENS, Universit\`{a} di Firenze, Via Nello Carrara
1, 50019 Sesto Fiorentino, Italy} \affiliation{\ Department of
Physics, Tsinghua University, Beijing 100084, China}
\author{Michele Modugno}
\affiliation{\ LENS - Dipartimento di Fisica, Universit\`{a} di
Firenze and INFM, Via  Nello Carrara 1, 50019 Sesto
Fiorentino,Italy} \affiliation{BEC-INFM Center, Universit\`a di
Trento, 38050 Povo, Italy}
\author{Hui Hu}
\affiliation{\ Abdus Salam International Center for Theoretical
Physics, P. O. Box 586, Trieste 34100, Italy}
\date{\today}

\begin{abstract}
By using the self-consistent Hartree-Fock-Bogoliubov-Popov theory,
we present a detailed study of the mean-field stability of
spherically trapped Bose-Fermi mixtures at finite temperature. We
find that, by increasing the temperature, the critical particle
number of bosons (or fermions) and the critical attractive
Bose-Fermi scattering length increase, leading to a significant
stabilization of the mixture.
\end{abstract}
\maketitle

The recent experimental achievement in the trapping and cooling of
degenerate Bose-Fermi mixtures of alkali-metal atoms have
introduced interesting new instances of a quantum many-body system
\cite {hulet,salomon,ketterle1,inguscio1,ketterle2}. Similar to
the purely Bose gases such mixtures provide a unique opportunity
of investigating fundamental quantum phenomena. Up to date a
number of theoretical analyses of trapped Bose-Fermi mixtures have
been presented, addressing, for example, the static property
\cite{molmer}, the phase diagram \cite{nygaard}, the behavior of
low-lying collective excitations \cite
{japan1,capuzzi1,japan2,liu1,capuzzi2,capuzzi3,liu2}, the
expansion \cite {hu1}, as well as the relevant implications for
the achievement of Bardeen-Cooper-Schrieffer (BCS) transition to a
superfluid phase of the fermionic components
\cite{stoof,heiselberg,viverit}. The last point is particularly
interesting and is highly favorable by tuning the Bose-Fermi
interaction by means of Feshbach resonance \cite{houbiers,simoni},
in order to induce a large effective attraction between fermionic
components by exchanging density fluctuations of the bosonic
background \cite {stoof,heiselberg,viverit}. It has been suggested
in Ref. \cite{simoni} that the optimal conditions for
boson-induced Cooper pairing are reached when the Bose-Fermi
mixtures are close to a critical point, beyond which the system
will become unstable and collapse.

This kind of collapse is indeed observed very recently in a binary
$^{40}\mathrm{K-}^{87}\mathrm{Rb}$ mixture \cite{inguscio2}. The
observation indicates a \emph{large} attractive interaction
between bosonic and fermionic atoms, and therefore highlights the
possibility of inducing a boson-mediated effective attraction
between two fermionic components once another hyperfine state of
fermionic atoms is populated. On the theoretical side, the
collapse or the instability of a trapped Bose-Fermi mixture at
\emph{zero} temperature has already been considered by several
authors \cite{molmer,japan3,roth1,roth2,modugno}. In Refs.
\cite{roth1,roth2}, the mean-field instability of a spherically
trapped Bose-Fermi mixture is studied using a standard two-fluid
model for the condensate and degenerate Fermi gas. Most recently,
the same mean-field model has been used in Ref. \cite{modugno}, in
which the precise geometry of the experiment is taken into
account. Its prediction has been compared with the experimental
stability diagram of $^{40}\mathrm{K}$-$^{87}\mathrm{Rb}$ mixtures
to give a better estimate of the $s$ -wave Bose-Fermi scattering
length \cite{modugno}.

In the present paper we investigate the effects of the temperature
on the stability of a spherically trapped Bose-Fermi mixture by
only a mean-field approach. The motivation is twofold: At first,
in the experiment there is a detection limit for determining the
temperature. This limit is around $0.67T_c$, where $T_c$ is the
transition temperature of the Bose-Einstein condensation (BEC).
Below this characteristic temperature, though no uncondensed
fraction of the BEC is practically detectable, the possible role
played by the thermal cloud of bosons in the stability of the
mixture should be clarified. On the other hand, in the context of
boson-induced BCS superfluidity, the critical temperature for the
formation of $s$-wave Cooper pairing increases exponentially as
the mixtures move to the collapse point. The qualitative or
quantitative understanding of the collapse of Bose-Fermi mixtures
at finite temperature is therefore crucial to optimize the
experimental conditions to achieve the BCS transition.

Our investigation is based on the simplest self--consistent
mean-field theory --- the Popov version of the
Hartree-Fock-Bogoliubov (HFB) theory --- that has been generalized
by us to the system of Bose-Fermi mixtures in previous works
\cite{hu2,liu2}. For a weakly interacting Bose gas, it was shown
that the HFB-Popov theory gives the correct thermodynamic
properties with a very good accuracy \cite{giorgini}, though it
fails to predict the correct behavior of collective excitations at
high temperatures \cite{ghfb}. In Ref. \cite{davis}, this theory
has been applied to study the collapse of attractive Bose-Einstein
condensates.

In the following we briefly summarize the main points of HFB-Popov
theory. The trapped binary mixture is considered as a
thermodynamic equilibrium system under the grand canonical
ensemble whose thermodynamic variables are $N_b$ and $N_f$,
respectively, the total number of trapped bosonic and fermionic
atoms, $T$, the absolute temperature, and $\mu _b$ and $\mu _f$,
the chemical potentials. In the second quantization language, the
density Hamiltonian of the mixture reads($\hbar =1$),
\begin{eqnarray}
\mathcal{H} &=&\mathcal{H}_b+\mathcal{H}_f+\mathcal{H}_{bf},  \nonumber \\
\mathcal{H}_b &=&\psi ^{+}(\mathbf{r})\left[
-\frac{\mathbf{\bigtriangledown
}^2}{2m_b}+V_{ext}^b(\mathbf{r})-\mu _b\right] \psi (\mathbf{r})+\frac{g_{bb}%
}2\psi ^{+}\psi ^{+}\psi \psi ,  \nonumber \\
\mathcal{H}_f &=&\phi ^{+}(\mathbf{r})\left[
-\frac{\mathbf{\bigtriangledown
}^2}{2m_f}+V_{ext}^f(\mathbf{r})-\mu _f\right] \phi (\mathbf{r}),
\nonumber
\\
\mathcal{H}_{bf} &=&g_{bf}\psi ^{+}(\mathbf{r})\psi (\mathbf{r})\phi ^{+}(%
\mathbf{r})\phi (\mathbf{r}), \label{hami}
\end{eqnarray}
where $\psi (\mathbf{r})$ ($\phi (\mathbf{r})$) is the Bose
(Fermi) field operator that annihilates an atom at position
$\mathbf{r}$. The first and second terms in the brackets contain
the kinetic-energy
operators and the external trapping potentials $V_{ext}^b(\mathbf{r}%
)=m_b\omega _b^2r^2/2$ and $V_{ext}^f(\mathbf{r})=m_f\omega
_f^2r^2/2$ for the bosonic and the fermionic species with masses
$m_b$ and $m_f$, respectively. In the dilute regime, we have
considered two types of contact interactions: the interactions
between bosons, and the interactions between
bosons and fermions. They are parametrized by the coupling constants $%
g_{bb}=4\pi \hbar ^2a_{bb}/m_b$ and $g_{bf}=2\pi \hbar
^2a_{bf}/m_r$,
respectively, in terms of the $s$-wave scattering length $a_{bb}$ and $%
a_{bf} $, with $m_r=m_bm_f/(m_b+m_f)$ being the reduced mass. We
have neglected here the fermion-fermion interactions since for
spin-polarized fermions the $s$-wave contact interaction is
prohibited by the Pauli principle. The next leading order,
$p$-wave interaction is small at low energy \cite {rokhsar} and
will not be considered in the following.

We consider the many-body ground state $\left| \Psi \right\rangle
$ of Hamiltonian (\ref
{hami}) as a direct product of a symmetric $N_b$-body state $%
\left| \Psi _b\right\rangle $ for the bosonic species and an antisymmetric $%
N_f$-body state $\left| \Psi _f\right\rangle $ for the fermions.
With this choice we are not considering the possible correlation
effects beyond the mean-field approximation \cite{note}. The
density Hamiltonian describing the Bose-Fermi coupling can
therefore be decoupled in a self-consistent mean-field manner,
namely,
\begin{equation}
\mathcal{H}_{bf}\simeq g_{bf}\left[ \psi ^{+}\psi \langle \phi
^{+}\phi \rangle +\langle \psi ^{+}\psi \rangle \phi ^{+}\phi
-\langle \psi ^{+}\psi \rangle \langle \phi ^{+}\phi \rangle
\right] .  \label{hbf}
\end{equation}
As a result, the bosonic and fermionic atoms experience,
respectively, the
effective potentials $V_{eff}^b(\mathbf{r})=V_{ext}^b(\mathbf{r})+g_{bf}n_f(%
\mathbf{r)}$ and $V_{eff}^f(\mathbf{r})=V_{ext}^f(\mathbf{r})+g_{bf}n_b(%
\mathbf{r),}$ with $n_b(\mathbf{r)}$ and $n_f(\mathbf{r)}$ being
the bosonic and fermionic density distributions, respectively. It
is then straightforward to apply the HFB-Popov theory for the
bosonic species, following Ref. \cite{griffin}. After invoking the
Bose symmetry breaking in the equation of motion for the Bose
field operator $\psi (\mathbf{r})$, we obtain, respectively, the
modified
Gross-Pitaevskii (GP) equation for the condensate wave function $\Phi (%
\mathbf{r}),$
\begin{equation}
\mathcal{L}_{GP}\Phi (\mathbf{r})=0,  \label{GPE}
\end{equation}
and the modified Bogoliubov-deGennes (BdG) equations for the
thermal quasiparticle amplitudes $u_i(\mathbf{r})$ and
$v_i(\mathbf{r})$,
\begin{eqnarray}
\left[ \mathcal{L}_{GP}+g_{bb}n_c(\mathbf{r})\right] u_i(\mathbf{r}%
)+g_{bb}n_c(\mathbf{r})v_i(\mathbf{r}) &=&\epsilon
_iu_i(\mathbf{r}),
\nonumber \\
\left[ \mathcal{L}_{GP}+g_{bb}n_c(\mathbf{r})\right] v_i(\mathbf{r}%
)+g_{bb}n_c(\mathbf{r})u_i(\mathbf{r}) &=&-\epsilon
_iv_i(\mathbf{r}), \label{BdG}
\end{eqnarray}
where the operator
\begin{equation}
\mathcal{L}_{GP}=-\frac{\mathbf{\bigtriangledown
}^2}{2m_b}+V_{ext}^b-\mu
_b+g_{bb}(n_c(\mathbf{r})+2n_T(\mathbf{r}))+g_{bf}n_f(\mathbf{r}),
\label{Lgp}
\end{equation}
is the generalized GP Hamiltonian that includes the mean-field
contributions generated by the interaction with the thermal cloud
and the fermionic species. Once these wave functions have been
determined, the local density of the condensate and of the
noncondensate, and the total bosonic density distribution can be
calculated according to
\begin{eqnarray}
n_c(\mathbf{r}) &=&\left| \Phi (\mathbf{r})\right| ^2,  \nonumber \\
n_T(\mathbf{r}) &=&\sum\limits_i\left[ \frac{\left( \left| u_i(\mathbf{%
r})\right| ^2+\left| v_i(\mathbf{r})\right| ^2\right) }{e^{\beta
\epsilon _i}-1}+\left| v_i(\mathbf{r})\right| ^2\right] ,
\nonumber \\  \label{nb}
 n_b(\mathbf{r)}
&=&n_c(\mathbf{r})+n_T(\mathbf{r}),
\end{eqnarray}
with $\beta =1/k_BT$ being the inverse temperature. To solve the
modified GP and BdG equations, one has to evaluate the fermionic
density distribution $n_f(\mathbf{r})$. To this aim, we employ the
finite-temperature Thomas-Fermi approximation (TFA)
\cite{rokhsar},
\begin{eqnarray}
n_f(\mathbf{r}) &=&\int \frac{d^3\mathbf{p}}{\left( 2\pi \right)
^3}\frac 1{e^{\beta \left(
\frac{\mathbf{p}^2}{2m_f}+V_{ext}^f+g_{bf}n_b-\mu
_f\right) }+1},  \nonumber \\
&=& \frac{m_f^{3/2}}{\sqrt{2}\pi^2} \int\limits_{-\infty
}^{+\infty }d\epsilon \frac 1{e^{\beta \left( \epsilon -\mu
_f\right)}+1}\left( \epsilon -V_{ext}^f-g_{bf}n_b\right) ^{1/2}.
\label{nf}
\end{eqnarray}
Specially, for the range of $N_f$ considered here
($N_f\simeq10^4$), it was shown that TFA is a good approximation
for a mixture with repulsive Bose-Fermi interaction at zero
temperature \cite{nygaard}. In our case, this approximation is
expected to work even better, since the quantum fluctuations are
suppressed by the finite temperature and the increased density due
to the Bose-Fermi attraction.

Equations (\ref{GPE})-(\ref{nf}) form a closed set of equations
that we have referred to as the ``HFB-Popov'' equations for a
dilute Bose-Fermi mixture. The simultaneous solution of this set
of equations gives the temperature-dependent density profiles of
the condensate, of the noncondensate, and of the degenerate Fermi
gas, from which we can extract the stability conditions of the
system. We have numerically solved these coupled equations by an
iterative scheme, as described in Ref. \cite{hu2}.

\begin{figure*}[tbp]
\centerline{\includegraphics[width=14.0cm,clip=]{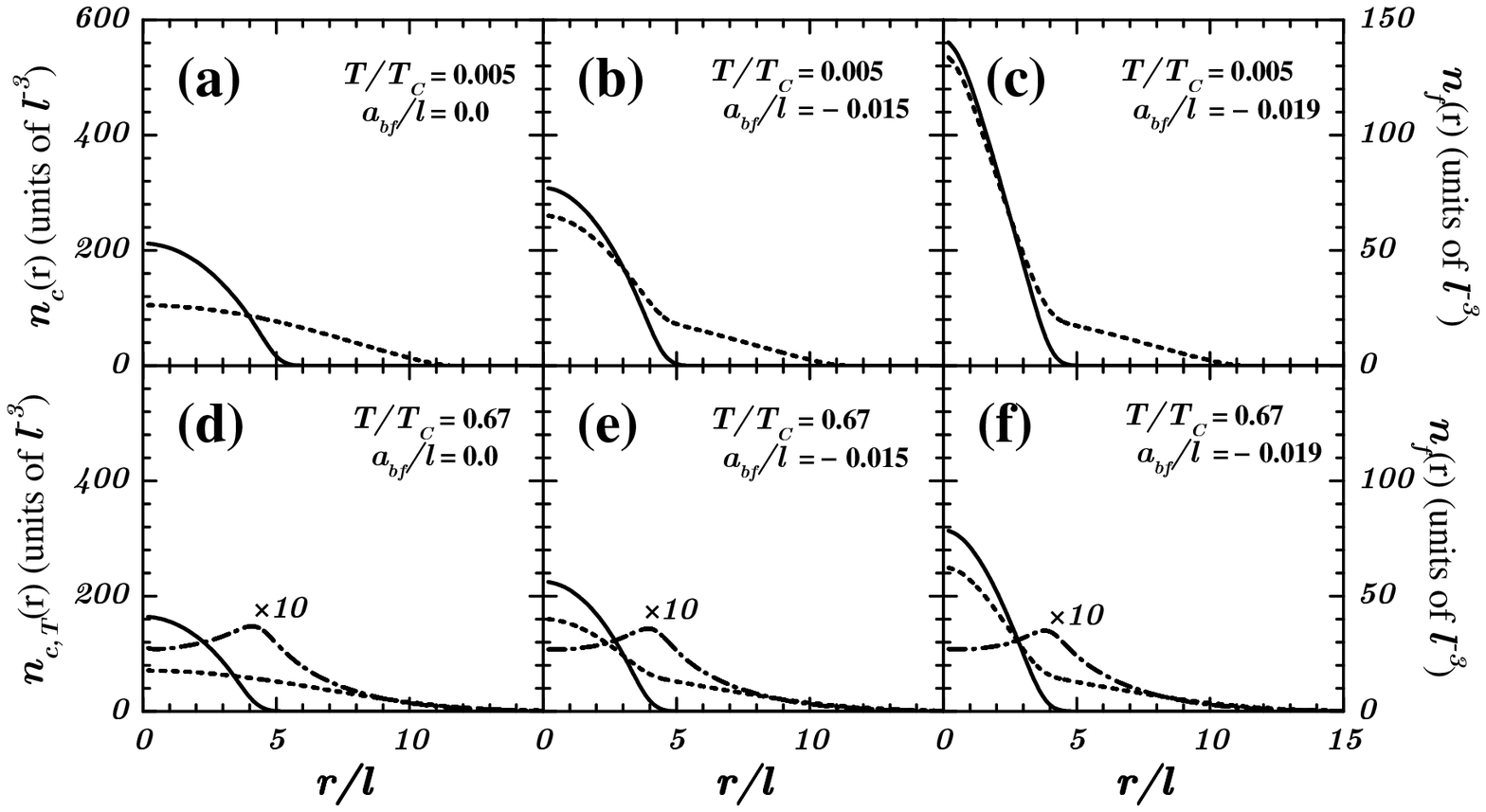}}
\caption{Radial density profiles of a Bose-Fermi mixture with $%
N_b=N_f=5\times 10^4$ for different temperatures and Bose-Fermi
interaction strengths. The condensate and noncondensate densities
$n_c(r)$ and $n_T(r)$ are given, respectively, by the solid and
dash-dotted lines (left scale), and the fermion density $n_f(r)$
by dashed line (right scale). To visualize $n_T(r)$ is enlarged by
a factor of ten. The upper and lower
rows show, respectively, examples with decreasing Bose-Fermi attraction at $%
T/T_C=0.005$ and $0.67$, where $T_C=0.94\hbar \omega%
N_b^{1/3}/k_B$ is the transition temperature of a non-interacting
Bose-Einstein condensate in the thermodynamic limit. We fix the
boson-boson scattering length $a_{bb}/l$ $=0.005$.} \label{fig1}
\end{figure*}

With above tools we investigate the effects of the temperature
upon the collapse of a trapped Bose-Fermi mixture induced by the
attractive Bose-Fermi interaction. Rather than taking the
realistic experimental conditions, in this paper we present a
general analysis of finite temperature effects, and restrict
ourselves to spherical symmetric systems. The reason is that for
this geometry the temperature-dependent density profiles of the
system are easily calculated. On the contrary, for anisotropic
systems, though the formalism given in the present paper is suited
as well, the numerical calculation is computationally much more
involved. In particular, for the highly anisotropic traps used in
the experiments at the European Laboratory for Non-linear
Spectroscopy (LENS) \cite{inguscio1,inguscio2} it is unlikely to
obtain useful information without the use of further
approximation. In the following, we consider equal masses for the
two species $m=m_b=m_f$ and identical trapping frequencies $\omega
=\omega _b=\omega _f$. The
oscillator length $l=\left( \hbar /m\omega \right) ^{1/2}$ and trap energy $%
\hbar \omega $, respectively, serve as fundamental length unit and
energy unit. We also take the transition temperature of a
non-interacting Bose gas with $N_b=5\times 10^4$ atoms,
$T_c=0.94\hbar \omega N_b^{1/3}/k_B=34.63\hbar \omega /k_B$, as a
characteristic temperature. In most calculations, we have set
$T=0.67T_c$ that corresponds to a typical detection limit for the
measurement of the temperature.

First we consider the effects of the temperature upon the density
profiles of the mixture. The upper and lower rows in Fig. 1 show
the density profiles of configurations with $N_b=N_f=5\times 10^4$
particles for three different values of the $s$-wave Bose-Fermi
scattering length $a_{bf}/l$ at two
temperatures: $T=0.005T_c$ and $T=0.67T_c$. The value of $a_{bb}/l=0.005$ %
has been fixed, which corresponds to $a_{bb}\approx%
100a_{\text{Bohr}}$ for a typical trap with $l=1\mu m$. Our
results at the lower temperature $T=0.005T_c$ are in good
agreement with the findings by Roth and co-workers
\cite{roth1,roth2}, that is, the mutual attraction between bosons
and fermions results in an enhancement of both densities in the
overlap region. In particular, as shown in Fig. (1c), both the
bosonic and fermionic densities grow substantially when the
mixture is close to the instability point (note that at
$T=0.005T_c$ the critical value of $a_{bf}/l$ is $-0.0193$). This
enhancement, however, is much reduced at a finite temperature,
whose effect is a broadening of the density distributions of the
condensate and of the Fermi gas and therefore reduces their center
densities. As can be seen in Fig. (1f), at $T=0.67T_c$ both
densities is decreased by a factor of 2 compared to the lower
temperature case. As a consequence, the mixture in this case is
expected to be much stabilized against collapse.

\begin{figure}[tbp]
\centerline{\includegraphics[width=7.5cm,clip=]{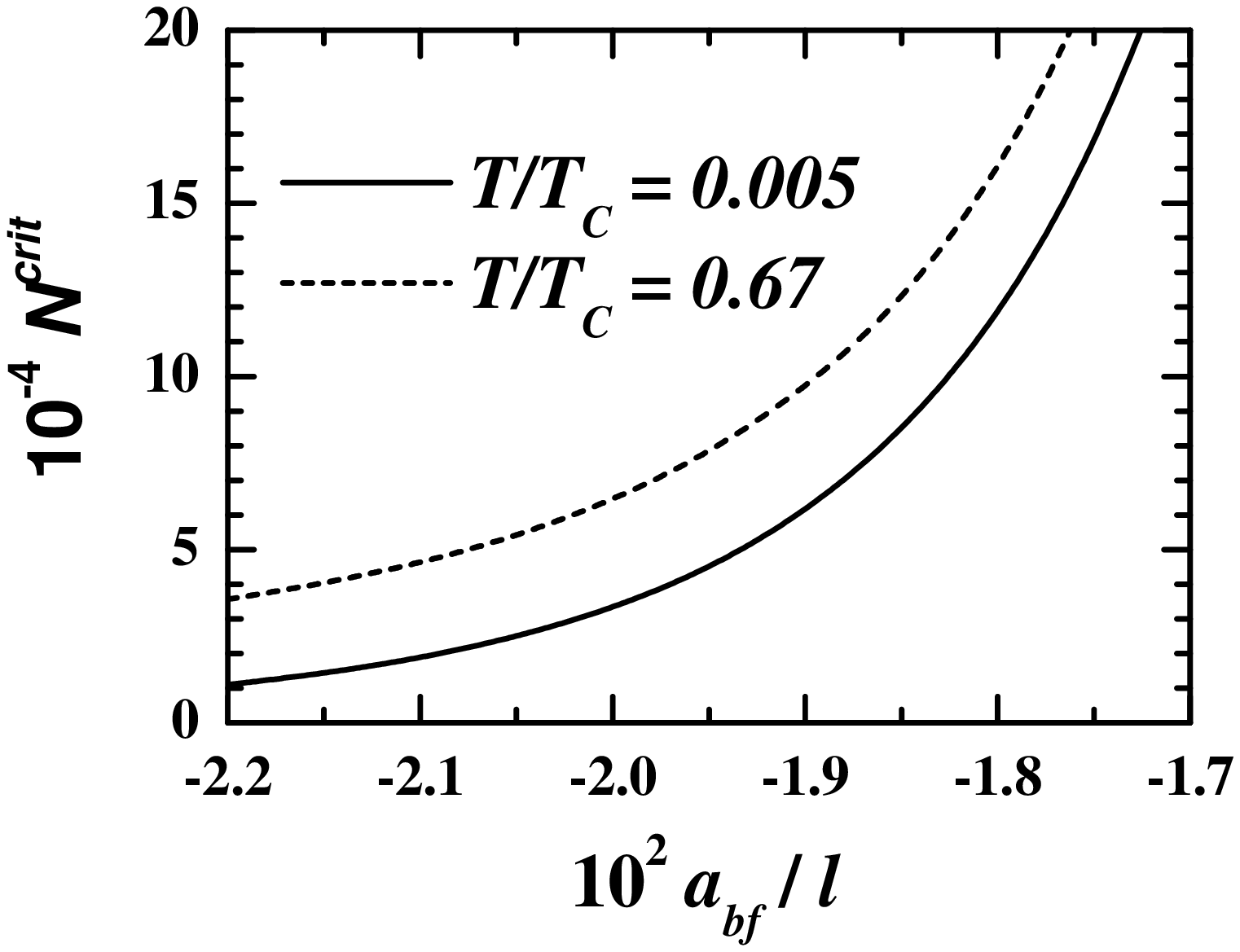}}
\caption{Behavior of the critical particle number as a function of
the $s$-wave Bose-Fermi scattering length $a_{bf}/l$ for $N_b=N_f$
at two temperatures: $T/T_C=0.005$ (solid line) and $T/T_C=0.67$
(dashed line). $T_C=0.94\hbar\omega N_b^{1/3}/k_B$ is the ideal
transition temperature evaluated at $N_b=5\times 10^4$. The
boson-boson scattering length $a_{bb}/l$ $=0.005$ is fixed.}
\label{fig2}
\end{figure}

In a simplified model \cite{molmer}, the collapse or the
instability of the mixture is governed by the balance between the
kinetic energy of fermions and the mutually attractive mean-field
generated by the Bose-Fermi interaction (see, for example, the
discussion above the Eq. (11) in Ref. \cite{molmer}), If the
Bose-Fermi attraction becomes too strong, \emph{i.e.}, the numbers
of bosons (or fermions) or the scattering length $a_{bf}$ becomes
too large, the attractive mean field cannot be stabilized by the
kinetic energy anymore, so both density distributions grow
indefinitely within the overlap region and collapse. According to
the study reported in Refs. \cite {molmer,roth1,roth2,modugno},
the onset of this mean-field instability is monitored by the
failure of the convergence during the iterative procedure. In this
manner, we can determine the critical $s$-wave Bose-Fermi
scattering length $a_{bf}$ or critical particle numbers
$N^{crit}_b$ and $N^{crit}_f$ beyond which the collapse occurs.

\begin{figure}[tbp]
\centerline{\includegraphics[width=7.0cm,clip=]{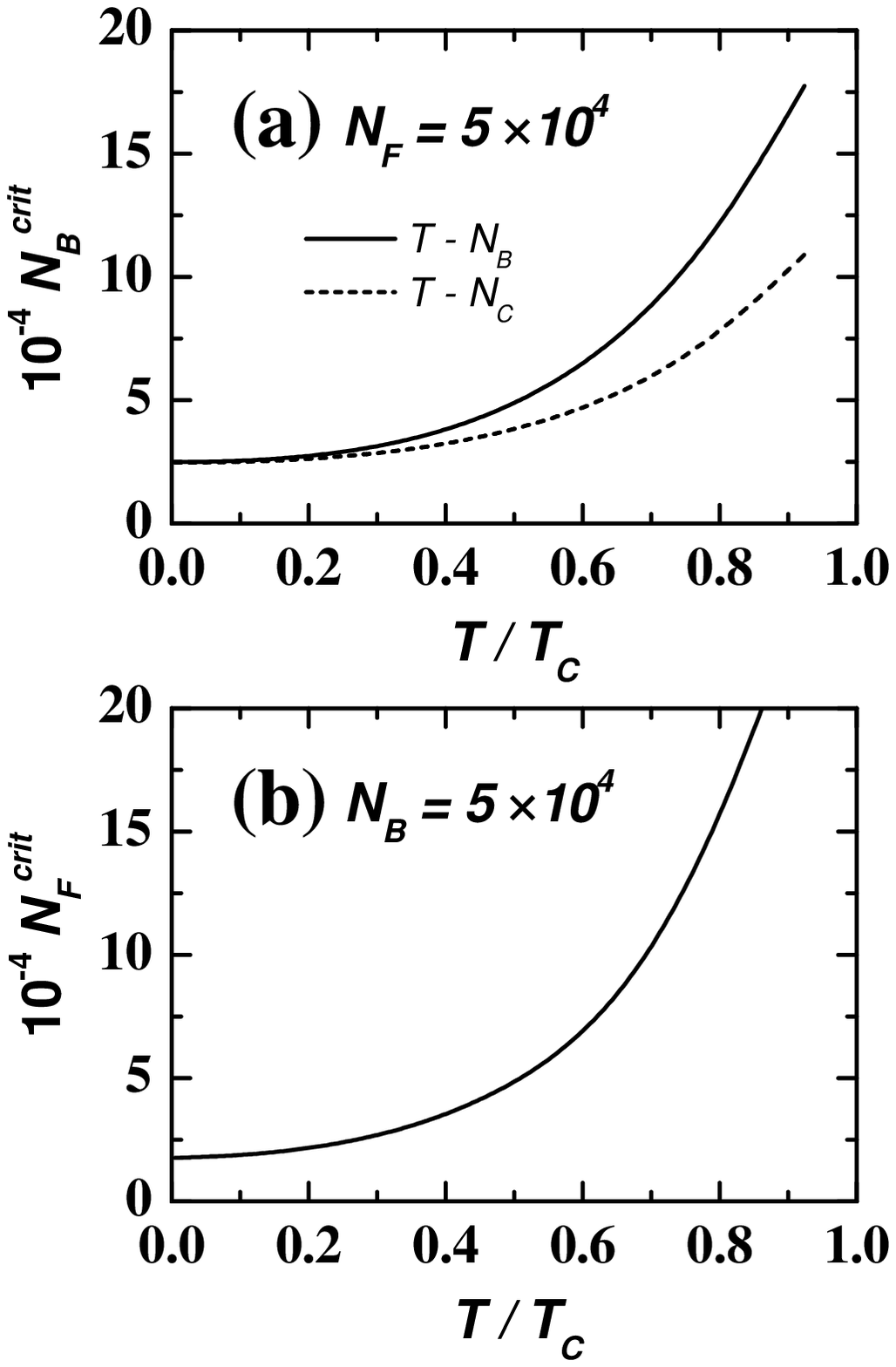}}
\caption{(a) The critical number of bosons $N_b^{crit}$ as a
function of the rescaled temperature $T/T_C$ with fixed number of
fermions $N_f=5\times 10^4$. The dashed line shows the
corresponding number of condensed bosons. (b) The critical number
of fermions $N_f^{crit}$ as a function of the re-scaled
temperature $T/T_C$ with fixed number of bosons $N_b=5\times%
10^4$. In both cases $T_C=0.94\hbar \omega N_b^{1/3}/k_B$ is the
ideal transition temperature for $N_b=5\times 10^4$. The other
parameters are: $a_{bb}/l$ $=0.005$ and $a_{bf}/l$ $=-0.020$. }
\label{fig3}
\end{figure}

In Fig. 2, we show the critical particle number as a function of
the $s$-wave Bose-Fermi scattering length $a_{bf}/l$ with fixed
$a_{bb}/l=0.005$ for two temperatures. In this calculation we have
kept equal numbers of bosons and fermions ($N_b=N_f$). At the
lower temperature $T=0.005T_c$, we observed that even a small
decrease of Bose-Fermi attraction can reduce the critical particle
number significantly, \emph{i.e.}, $N^{crit}$ changes by a factor
of 2 when $a_{bf}/l$ changes by 5 percent. This strong dependence
of the criticality on $a_{bf}/l$ is in accordance with the scaling
law for the critical particle number: $N^{crit}\sim
|a_{bf}|^{-\alpha}$ with $\alpha=12$, as discussed in Refs.
\cite{molmer,japan3,modugno}. The presence of a moderate
temperature $T=0.67T_c$ results in a substantial increase of the
critical particle number. For example, at $a_{bf}/l=-0.022$,
$N^{crit}$ grows from $1.1\times 10^4$ to $3.6\times 10^4$ with
the inclusion of the temperature. Apart from this growth, the
critical particle number is still strongly dependent on
$a_{bf}/l$. However, the presence of the finite temperature leads
to a re-normalization of the scaling exponent, \emph{i.e.},
$\alpha\sim8$.

In Figs. (3a) and (3b), we study the critical particle numbers of
bosons and fermions as a function of the temperature. We report,
respectively, the prediction on the critical particle numbers of
bosons (with fixed $N_f=5\times 10^4$) and fermions (with fixed
$N_b=5\times 10^4$). For comparison, in Fig. (3a), we also show
the critical number of the condensed atoms against
temperature(dashed line). As expected, the critical particle
number for each species increase with increasing the temperature.
The dependence is highly nonlinear. Below $0.5T_c$, the critical
particle number varies slowly with the temperature, whereas above
$0.5T_c$ it rises up steeply. In addition, the critical particle
number of fermions grows more rapidly than that of bosons against
the temperature.

\begin{figure}[tbp]
\centerline{\includegraphics[width=7.5cm,clip=]{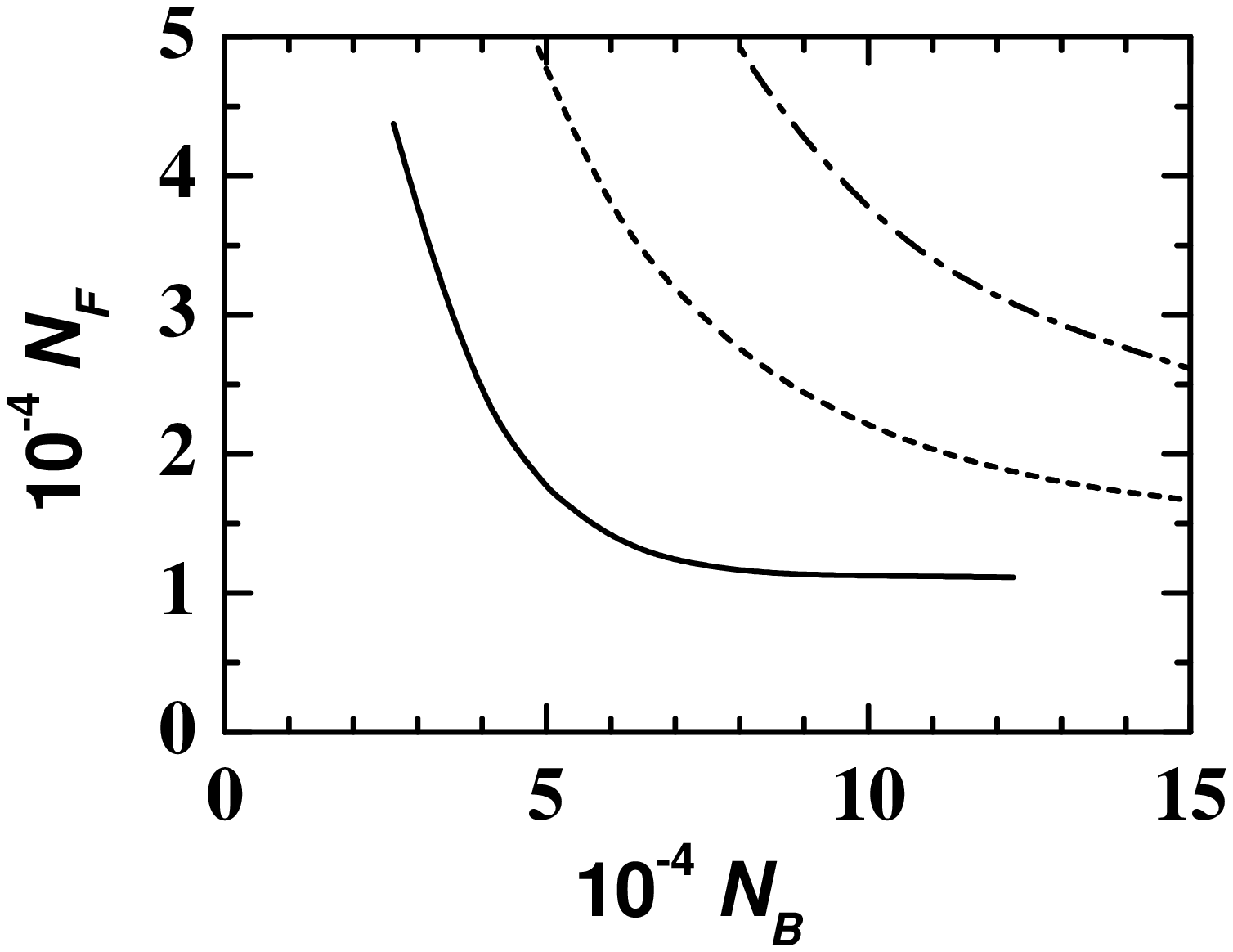}}
\caption{Region of stability of the Bose-Fermi mixture, as a
function of the
number of atoms, for the case of $a_{bb}/l$ $=0.005$ and $a_{bf}/l$ $=-0.020$%
. Lines show the boundary between the stable (left) and collapse (right)
regions, for three values of the temperature: $T/T_C=0.005$ (solid line), $%
T/T_C=0.50$ (dashed line) and $T/T_C=0.67$ (dash-dotted line), where $%
T_C=0.94\hbar \omega N_b^{1/3}/k_B$ is the transition temperature
for a non-interacting Bose gas with $N_b=5\times 10^4$.}
\label{fig4}
\end{figure}

Finally, we have built a region of stability of the Bose-Fermi
mixture in Fig. 4, as a function of the number of atoms, for the
cases of $a_{bb}/l=0.005$ and $a_{bf}/l=-0.020$, for three values
of the temperature: $T/T_C=0.005$, $T/T_C=0.50$ and $T/T_C=0.67$.
Each of the curves marks the limit of stability. For numbers of
bosons or fermions below the stability limit the mixture is
stable, otherwise the mixture is unstable against mean-field
collapse. Figure 4 indicates that the region of the stability
broadens with increasing temperature. This behavior emphasize
again that the inclusion of a temperature gives rise to a
significant stabilization of the mixtures.

In conclusion, we have investigated the mean-field stability of a
spherically trapped binary Bose-Fermi mixture at finite
temperature. We solved the coupled HFB-Popov equations numerically
and obtained the critical particle number as a function of the
temperature and the $s$-wave Bose-Fermi scattering length. We have
shown that the critical particle number and the critical
Bose-Fermi scattering length increase with the inclusion of a
moderate temperature that corresponds to the typical experimental
detection limit. This leads to a significant stabilization of the
Bose-Fermi mixtures.

\begin{acknowledgments}
One of the authors (X.-J.L) was supported by the NSF-China under
Grant No. 10205022.
\end{acknowledgments}

\end{document}